\documentclass[12pt]{article}
\usepackage[latin1]{inputenc}
\usepackage{amsmath}
\usepackage{amsfonts}
\usepackage{amssymb}
\usepackage{hyperref}
\usepackage{psfrag}
\usepackage[dvips]{graphicx}
\usepackage{colordvi}
\usepackage{times}
\usepackage[margin=1cm,nohead]{geometry}

\newcommand{\bc}{\begin{center}}
\newcommand{\ec}{\end{center}}
\newcommand{\be}{\begin{equation}}
\newcommand{\ee}{\end{equation}}
\newcommand{\bna}{\begin{eqnarray}}
\newcommand{\ena}{\end{eqnarray}}

\newcommand{\mpaa}{\begin{minipage}[t]{6cm}}
\newcommand{\mpea}{\end{minipage}}
\newcommand{\mpab}{\begin{minipage}[t]{8cm}}
\newcommand{\mpeb}{\end{minipage}}
\newcommand{\mpac}{\begin{minipage}[t]{13cm}}
\newcommand{\mpec}{\end{minipage}}
\newcommand{\mpad}{\begin{minipage}[t]{13cm}}
\newcommand{\mped}{\end{minipage}}
\newcommand{\mpae}{\begin{minipage}[t]{13cm}}
\newcommand{\mpee}{\end{minipage}}
\newcommand{\mpaf}{\begin{minipage}[t]{6cm}}
\newcommand{\mpef}{\end{minipage}}

\newcommand {\bdm} {\begin{displaymath}}
\newcommand {\edm} {\end{displaymath}}

\usepackage{color}
\definecolor{darkblue}{rgb}{0,0,0.6}
\definecolor{darkred}{rgb}{0.7,0,0}
\definecolor{darkgreen}{rgb}{0,0.7,0}
\usepackage{amsfonts}

\newcommand{\bea}{\begin{eqnarray}}
\newcommand{\eea}{\end{eqnarray}}

\def\XXint#1#2#3{{\setbox0=\hbox{$#1{#2#3}{\int}$}
     \vcenter{\hbox{$#2#3$}}\kern-.5\wd0}}

\begin{document}

\title{Beyond the linear Fluctuation-Dissipation Theorem: the Role of Causality.}
\author{Valerio Lucarini$^{1,2}$ [valerio.lucarini@zmaw.de]\\
$^1$Meteorologisches Institut, University of Hamburg\\KlimaCampus, Grindelberg 7,
20144 Hamburg, Germany\\
$^2$Department of Mathematics and Statistics\\University of Reading, Reading, RG6
6AX, UK\\ 
\\
Matteo Colangeli$^3$\\
$^3$Dipartimento di Matematica, Politecnico di Torino\\Corso Duca degli Abruzzi 24,
10129 Torino, Italy}

\maketitle

\abstract{In this paper we re-examine the traditional problem of connecting the internal fluctuations of a system to its response to external forcings and extend the classical theory in order to be able to encompass also nonlinear processes. With this goal, we try to join on the results by Kubo on statistical mechanical systems close to equilibrium, i.e. whose unperturbed state can be described by a canonical ensemble, the theory of dispersion relations, and the response theory recently developed by Ruelle for non-equilibrium systems equipped with an invariant SRB measure. Our derivations highlight the strong link between causality and the possibility of connecting unambiguously fluctuation and response, both at linear and nonlinear level. We first show in a rather general setting how the formalism of the Ruelle response theory can be used to derive in a novel way Kramers-Kronig relations connecting the real and imaginary part of the linear and nonlinear response to external perturbations. We then provide a formal extension at each order of nonlinearity of the fluctuation-dissipation theorem (FDT) for general systems possessing a smooth invariant measure. Finally, we focus on the physically relevant case of systems close to equilibrium, for which we present explicit fluctuation-dissipation
relations linking the susceptibility describing the $n^{th}$ order response of the system with the expectation value of suitably
defined correlations of $n+1$ observables taken in the equilibrium ensemble. While the FDT has an especially compact structure in the linear case, in the nonlinear case joining the statistical properties of the fluctuations of the system to its response to external perturbations requires
linear changes of variables, simple algebraic sums and multiplications, and a
multiple convolution integral. These operations, albeit cumbersome, can be easily implemented
numerically.}

\section{Introduction}

The equilibrium statistical mechanical description of a many-particle system is
rooted on the Hamiltonian equations of motion, which feature no dissipation, i.e. a
vanishing average phase space contraction rate. This basic tenet of the Hamiltonian
dynamics is mirrored, in the Gibbs' ensemble approach, by the Liouville Theorem,
which claims the conservation of the probability measure in the phase space. 

On the other hand, nonequilibrium statistical mechanics deals with the investigation
of dissipative dynamical systems attaining a steady state. 
Thus, the mathematical description is richer: beside the many-particle system, one
needs to suitably take into account the effects of the external field performing
work on the system, and of a thermal reservoir, which absorbs the heat generated
within the system by the action of the external field.
One of the few general results, in this setting, is represented by the Fluctuation
Relations (FR), which concern some peculiar symmetry properties of the underlying
microscopic dynamics under time-reversal, with respect to the phase space
contraction rate (which can be replaced, in stochastic dynamics, by the Entropy
production of the system).
The FR hold for systems arbitrarily far from equilibrium and may be essentially
regarded as a large deviation result \cite{Gallavotti, Colangeli}.

When considering weakly dissipative systems, the Fluctuation-Dissipation Theorem (FDT) \cite{Kubo,callen} establishes a link between the response properties of a system to an external
(possibly time-dependent) perturbation and a correlation function computed in the
equilibrium ensemble.  The idea underlying the FDT is simple, yet powerful: for linear deviations from
equilibrium, one may compute, for instance, the viscosity of a fluid or the
resistance of an electrical wire (which are related to the dissipation produced in
the system when an external driving, i.e. a shear stress or a voltage, is switched
on) without actually applying the external field, but just through observing suitably defined correlation properties of the fluctuations of the unperturbed system. In other terms, tit is possible to establish a correspondence between the internal fluctuations of a system and its response to weak external forcings. The practical as well as conceptual implications of this result are immense.  

The problem of determining the response of the system immersed in a thermal bath to
external perturbations was generalised by Kubo himself in order to study higher
order effects on observable, which become practically relevant in the presence of stronger forcings.  This line of investigation has since start found relevant applications in fields such as optics
where by the '60s the revolutionary laser technology allowed for studying intense
radiation-matter coupling and for generating a complex and fascinating phenomenology
with wide-ranging theoretical as well as industrial relevance. See the book by
Bloembergen \cite{bloembergen} for a interesting mix of early appraisal of Kubo's
work and fresh outlook on the very first studies in nonlinear optics, and the book
by Butcher and Cotter \cite{butcher} for a more recent point of view. For rather obvious
reasons, optics has historically been the scientific context where the response
theory has been developed extensively focusing on the frequency domain, rather than
on the time domain, description. This has led to emphasizing the link between the
fact that the system obeys causality when responding to an external perturbation and
the existence of general integral dispersion relations, commonly known as
Kramers-Kronig relations, linking the real and imaginary part of the
frequency-dependent response to radiation, and the related sum rules. This has been of great practical relevance for computing and reconstructing the optical properties of matter, natural as well as artificial. An extensive account of this line of work can be found
in \cite{peiponen99,lucarini05}.

Along a different route, the investigation of the response to external perturbations
has been extended to encompass the case where the unperturbed system is not  at
equilibrium, but is rather forced and dissipative, and lives in a non-equilibrium
steady state (NESS) \cite{galla06}. In this state, the phase space continuously contracts, entropy
is generated, and SRB measures \cite{eckmann85,ruelle89,young2002} provide the
natural mathematical framework for describing its statistical properties. 
Ruelle~\cite{ruelle98} recently derived explicit formulas for
describing the smooth dependence of the SRB measure to small perturbations of the
flow in the case of Axiom A systems \cite{young2002}. Such response theory boils down to a Kubo-like perturbative expression
connecting the terms describing the linear and nonlinear response of the system as
expectation values of observables on the unperturbed SRB measure. This approach is
especially useful for studying the impact of changes in the internal parameters of a
system or of small modulations to the external forcing, and various studies have
highlighted the practical relevance of Ruelle theory for studying what we may call
the sensitivity of the system to small perturbations. In some cases, the emphasis
has been on providing convincing ways to compute the linear response from the
unperturbed motion \cite{majda07}, in other studies, the authors have highlighted
the properties of the frequency dependent linear response of the system
\cite{reick02,lucarinisarno11}. Finally, some efforts have been directed at extending the
analysis of the frequency dependent response to the nonlinear case \cite{lucarini08}
and on testing the robustness of the theory with simple chaotic models
\cite{lucarini09}. Recently the response theory has been used to study the impact of stochastic perturbations \cite{lucarini12} and derive rigorously parametrizations for reducing the complexity of multiscale systems \cite{wouters12}, which provides potentially interesting links to the Mori-Zwanzig projection theory \cite{Zwanzig01}. 

The link between linear response of the system to external perturbations and its
internal fluctuations is more elusive when the unperturbed state is a NESS. In Refs.
\cite{ruelle98,Sepulchre} is is shown that since the invariant measure is singular,
the response of the system contains two contributions, such that the first may
be expressed in terms of a correlation function evaluated with respect to the
unperturbed dynamics along the space tangent to the attractor and represents the
dissipative version of the equilibrium correlation function occurring in Kubo's
theory \cite{ruelle09,lucarinisarno11}. On the other hand, the second term, which has no
equilibrium counterpart, depends on the dynamics along the stable manifold, and,
hence, it may not be determined from the unperturbed dynamics and is also quite
difficult to compute numerically. This expresses the basic fact, already suggested heuristically by Lorenz \cite{lorenz79}, that  in the case of non-equilibrium systems internal and forced fluctuations of the system are not equivalent.
When devising algorithms for computing the response of the system, in fact, Majda and
collaborators are forced to use different methods for computing the correlation-like
and the additional term described above \cite{majda07}. 
The relevance of the novel term spoiling the canonical structure of the FDT
for dissipative chaotic systems is still a matter of ongoing research. While Majda and collaborators find this term to be of comparable size as the one coming from the usual correlation integral, in Ref. \cite{ColRonVul} it is shown, by means of low-dimensional
solvable models, that the novel term introduced by Ruelle is expected to attain its
own relevance only in very peculiar situations, such as systems with carefully
oriented manifolds in phase space and for initial perturbations chosen along the
stable directions. These observations stem from the fact that physics is mainly
concerned with smooth observables and with projections from high-dimensional spaces
to lower dimensional ones \cite{EvRon,BPRV}. This may explain why some attempts of reconstructing the response to perturbations of a complex system such as the climate via the application of the classic FDT have enjoyed a good success \cite{alexeev,branstator,ring}, even if it is clear that the performance depends critically on the choice of the observable of interest. Moreover, it is important to underline that recent works \cite{vulpiani07} have emphasized that FDT applies for all systems whose invariant measure is smooth, which is, in particular, the case for deterministic systems perturbed with noise \cite{boffetta03}. 



The purpose of this paper is twofold: we emphasize the intimate link existing between the causality of
the response of the system and the possibility of connecting response and
fluctuations. First, we wish to show how the Ruelle response theory can be used to derive straightforwardly Kramers-Kronig relations connecting at all orders of nonlinearity the real and imaginary part of the susceptibility - frequency dependent response of the system to perturbations -, and how the susceptibility can be written in terms of unperturbed properties of the system. This is accomplished in Sec. \ref{sec:respfun}. Subsequently, we focus on extending the FDT.  In Sec. \ref{sec:fdt1} we derive a nonlinear generalization of the classical FDT by considering higher orders in the standard perturbative expansion around the unperturbed measure, assumed to be absolutely continuous with respect to Lebesgue. In Sec. \ref{sec:fdt2} we present explicit calculations for the canonical equilibrium reference frame, which provide the natural extension to arbitrary order of the classical Kubo's FDT. In Sec. \ref{sec:conc}  we present our conclusions and perspectives for future works. Finally, as a side note, in App. \ref{sec:app} we show that, when considering perturbations to a canonical ensemble, at all orders of nonlinearity the imaginary part of the susceptibility of the observable conjugated to the external field is intimately connected to dissipation.



\section{Response theory and Kramers-Kronig relations}
\label{sec:respfun}
Ruelle~\cite{ruelle98} recently derived explicit formulas for describing the smooth dependence of the SRB measure of Axiom A dynamical systems to small perturbations of the flow. Such response theory boils down to a Kubo-like perturbative expression connecting the terms describing the linear and nonlinear response of the system as expectation values of observables on the unperturbed measure. At order $n$ of nonlinearity, such expectation values can be written as $n-$uple convolution of a causal Green function with the time-delayed perturbative fields, so that at every order Kramers-Kronig relations can be written for the Fourier transform of the Green function, the so-called susceptibility ~\cite{lucarini08}.

Let's consider a general dynamical system whose evolution equation can be written as $\dot{x} = F(x)$ and let's assume that it possesses an invariant SRB measure $\rho^{(0)}$. Ruelle \cite{ruelle97,ruelle09,lucarini08} has shown that if the system is weakly perturbed so that its evolution equation can be written as:
\begin{align}
\dot{x} = F(x)+X(x)T(t)
\end{align}
where $X(x)$ is a weak time-independent forcing and $T(t)$ is its time modulation, it is possible to write the modification to the expectation value of a general observable $A$ as a perturbative series:
\begin{equation}
\rho(A)_t=\sum_{j=0}^\infty \rho^{(n)}(A)_t,
\end{equation}
where $\rho^{(0)}$ is the unpertubed invariant measure, $\rho^{(n)}(A)_t$ with $n\geq1$ represents the contribution due to $n^{th}$ order nonlinear processes and can be expressed as  a $n-$uple convolution product: 
\begin{equation}
\rho^{(n)}(A)_t,=\int_{-\infty}^\infty d\tau_1\ldots \int_{-\infty}^\infty d\tau_n G(\tau_1,\ldots,\tau_n)T(t-\tau_1)T(t-\tau_n)\label{deltan}.
\end{equation}
The integration kernel $G^{(n)}(\tau_1,\ldots,\tau_n)$ is the $n^{th}$ order Green function, which can be written as:
\begin{align}
G(\tau_1,\ldots,\tau_n)&=\int \rho^{(0)}(dx) \Theta(\tau_1)\Theta(\tau_2-\tau_1)\ldots \Theta(\tau_n-\tau_{n-1})\Lambda \Pi(\tau_n-\tau_{n-1}) \Lambda \Pi(\tau_{n-1}-\tau_{n-2}) \Lambda \Pi(\tau_{1})A(x)
 \label{Gn}.
\end{align}
where $\Lambda(\bullet)= X\cdot \nabla(\bullet)$ describes the impact of the perturbation field and $\Pi(\sigma)$ is the unperturbed time evolution operator such that $\Pi(\sigma)K(x)= K(x(\sigma))$. The Green function obeys two fundamental properties
\begin{itemize}
\item its variables are time-ordered: if $j>k$, $\tau_j>\tau_k\rightarrow G^{(n)}(\tau_1,\ldots,\tau_n)=0$;
\item the function is causal: $\tau_1<0\rightarrow G^{(n)}(\tau_1,\ldots,\tau_n)=0$. 
\end{itemize}
Obviously, in the linear case only the second condition applies. 
These properties
allow rewriting the Green function as
$G(\tau_1,\ldots,\tau_n)=\Theta(\tau_1)\Pi_{j=2}^{n}\Theta(\tau_j-\tau_{j-1})R(\tau_1,\ldots,\tau_n)$,
where the Heaviside distribution $\Theta(\tau_1)$ takes care of guaranteeing the
causality, the terms of the form $\Theta(\tau_j-\tau_{j-1})$ enforce the
time-ordering, while $R(\tau_1,\ldots, \tau_{n})$ is the Response Function, which
contains the information
about the microscopic dynamics of the system. 
By applying the Fourier transform to Eq. \ref{deltan}, with $\hat{Y}(\omega)=\mathcal{F}(Y(t))=\int_{-\infty}^\infty d t \exp[-i\omega t] Y(t)$ one obtains the following
expression \cite{lucarini08}:
\begin{equation}
\rho^{(n)}\hat{(A)}\langle(\omega)=\int_{-\infty}^\infty d\omega_1\ldots
\int_{-\infty}^\infty d\omega_n \chi^{(n)}(\omega_1,\ldots,\omega_n)
\hat{T}(\omega_1)\ldots
\hat{T}(\omega_n)\delta(\omega-\sum_{j=1}^{n}\omega_j)\label{deltanomega},
\end{equation}
where
$\hat{\rho^{(n)}(A)\langle}(\omega)=\mathcal{F}(\rho^{(n)}(A)_t)$,
$\hat{T}(\omega_j)=\mathcal{F}(T(\tau_j))$, and the susceptibility
$\chi^{(n)}(\omega_1,\ldots,\omega_n)$ is the $n-$dimensional Fourier transform of
$G^{(n)}(\tau_1,\ldots,\tau_n)$ defined as :
\begin{align}
\chi^{(n)}(\omega_1,\ldots,\omega_n)&=\int_{-\infty}^\infty d\tau_1 \ldots d\tau_n
\exp[-i\omega_1 \tau_1]\ldots \exp[-i\omega_n
\tau_n]G^{(n)}(\tau_1,\tau_2,\ldots,\tau_n)
\label{chisusc},
\end{align}
while the term containing the Dirac $\delta$ ensures that the frequency of the
output is identical to the sum of the input frequencies. Note that we use the same definition for the Fourier transform as in \cite{Kubo}, while the sign of the frequency variable in the integration is opposite to what used in \cite{lucarini05,lucarini08}, which is more common in the optical literature. 
Defining: 
\begin{align}
G_{S}^{(n)}(\tau_1,\ldots,\tau_n)&=G^{(n)}(\tau_1,\ldots,\tau_n)+G^{(n)}(-\tau_1,\ldots,-\tau_n)\\
G_{A}^{(n)}(\tau_1,\ldots,\tau_n)&=G^{(n)}(\tau_1,\ldots,\tau_n)-G^{(n)}(-\tau_1,\ldots,-\tau_n)\label{gdef}
\end{align}
which are different from zero for $\tau_n>\tau_{n-1}>...>\tau_1>0$ and
$\tau_n<\tau_{n-1}<...<\tau_1<0$ and have opposite parity with respect to exchange
of the sign of all of the variables, being $G_{S}^{(n)}$ even and $G_{A}^{(n)}$ odd
with respect to this symmetry. We have that:
\begin{align}
2\Re\{\chi^{(n)}(\omega_1,\ldots,\omega_n)\}&=\int_{-\infty}^\infty d\tau_1\ldots 
d\tau_n \exp[-i\omega_1 \tau_1]\ldots \exp[-i\omega_n
\tau_n]G_S^{(n)}(\tau_1,\ldots,\tau_n)\\
2i\Im\{\chi^{(n)}(\omega_1,\ldots,\omega_n)\}&=\int_{-\infty}^\infty d\tau_1\ldots 
d\tau_n \exp[-i\omega_1 \tau_1]\ldots \exp[-i\omega_n
\tau_n]G_A^{(n)}(\tau_1,\ldots,\tau_n).\nonumber\\
\label{chisuscb}
\end{align}
Thanks to causality $\forall j$ we have that 
\begin{align}
G^{(n)}(\tau_1,\ldots,\tau_n)=\Theta(\tau_j)G^{(n)}(\tau_1,\ldots,\tau_n)\nonumber\\
\Theta(\tau_j)G_{S}^{(n)}(\tau_1,\ldots,\tau_n)\nonumber\\
\Theta(\tau_j)G_{A}^{(n)}(\tau_1,\ldots,\tau_n).
\label{causal}
\end{align}
By applying the Fourier Transform to these identities and using the convolution
theorem we obtain:
\begin{align}
\chi^{(n)}(\omega_1,\ldots,\omega_n)&
=\frac{1}{2\pi}\left(\hat{\Theta}(\omega_j)\right)*\left(2\Re\{\chi^{(n)}(\omega_1,\ldots,\omega_j,\ldots,\omega_n)\}\right)\nonumber\\
&
=\frac{1}{2\pi}\left(\hat{\Theta}(\omega_j)\right)*\left(2i\Im\{\chi^{(n)}(\omega_1,\ldots,\omega_j,\ldots,\omega_n)\}\right)\nonumber\\
&
=\frac{1}{2\pi}\left(\hat{\Theta}(\omega_j)\right)*\left(\chi^{(n)}(\omega_1,\ldots,\omega_j,\ldots,\omega_n)\right)\nonumber\\
&
=\left(-\frac{i}{\pi}\mathcal{P}\left(\frac{1}{\omega_j}\right)+\delta\left(\omega_j\right)\right)*\left(\Re\{\chi^{(n)}(\omega_1,\ldots,\omega_j,\ldots,\omega_n)\}\right)\nonumber\\
&
=\left(-\frac{i}{\pi}\mathcal{P}\left(\frac{1}{\omega_j}\right)+\delta\left(\omega_j\right)\right)*\left(i\Im\{\chi^{(n)}(\omega_1,\ldots,\omega_j,\ldots,\omega_n)\}\right)\nonumber\\
&
=\left(-\frac{i}{\pi}\mathcal{P}\left(\frac{1}{\omega_j}\right)+\delta\left(\omega_j\right)\right)*\left(\frac{1}{2}\chi^{(n)}(\omega_1,\ldots,\omega_j,\ldots,\omega_n)\right)
\end{align}
where $\mathcal{P}$ indicates that the integral must be computed considering the
principal part and * indicates the operation of convolution product.  As the same
causality argument given in Eq. \ref{causal} can be repeated for any time variable
$\tau_k$, we have that: 
\begin{align}
G^{(n)}(\tau_1,\ldots,\tau_n)=\Pi_{i=1}^k
\Theta(\tau_{j_i})G^{(n)}(\tau_1,\ldots,\tau_n)\label{causal2}\\
\Pi_{i=1}^k\Theta(\tau_{j_i})G_{S}^{(n)}(\tau_1,\ldots,\tau_n)\nonumber\\
\Pi_{i=1}^k\Theta(\tau_{j_i})G_{A}^{(n)}(\tau_1,\ldots,\tau_n) ,
\end{align}
where $j_1,\ldots,j_k$ runs over some or all of the indices $1,\ldots,n$. When
taking the Fourier Transform of the previous identities, we obtain: 
\begin{align}
\chi^{(n)}(\omega_1,\ldots,\omega_n)& =\Pi_{i=1}^k\left(-\frac{i}{\pi}\mathcal{P}\left(\frac{1}{\omega_{j_i}}\right)+\delta\left(\omega_{j_1}\right)\right)*\left(\frac{1}{2^{k-1}}\Re\{\chi^{(n)}(\omega_1,\ldots,\omega_j,\ldots,\omega_n)\}\right)\nonumber\\
& =\Pi_{i=1}^k\left(-\frac{i}{\pi}\mathcal{P}\left(\frac{1}{\omega_{j_i}}\right)+\delta\left(\omega_{j_1}\right)\right)*\left(\frac{i}{2^{k-1}}\Im\{\chi^{(n)}(\omega_1,\ldots,\omega_j,\ldots,\omega_n)\}\right)\nonumber\\
& =\Pi_{i=1}^k\left(-\frac{i}{\pi}\mathcal{P}\left(\frac{1}{\omega_{j_i}}\right)+\delta\left(\omega_{j_1}\right)\right)*\left(\frac{1}{2^k}\chi^{(n)}(\omega_1,\ldots,\omega_j,\ldots,\omega_n)\right)
\label{kkgen}
\end{align}
where * must be intended as multiple convolution product of the variables
$\omega_{j_1},\ldots,\omega_{j_k}$. Equations \ref{kkgen} provide an alternative
expression of generalized Kramers-Kronig relations for nonlinear susceptibilities,
first presented in the special case of optical processes in \cite{lucarini05}.  

We emphasize that Eqs. \ref{gdef}- \ref{kkgen} provide a fundamental connection
between the time-dependent response of the system to perturbations and the real and
imaginary part of the susceptibility. The Kramers-Kronig relations establish the
correspondence between the fundamental property of causality in the response with
the fact that the knowledge of only either the real or the imaginary part of the
susceptibility is sufficient to reconstruct the full frequency dependent response of
the system, both in the linear and in the nonlinear regime.  

\subsection{Response Function}
We now take a slightly different way for analyzing the frequency dependent response of the system by focusing on the  the Fourier transform of the response function $R(\tau_1,\ldots,\tau_n)$. We rewrite
the definition of the $n^{th}$ order susceptibility as follows: 
\begin{align}
\chi^{(n)}(\omega_1,\ldots,\omega_n)&=\int_{-\infty}^\infty d\tau_1 \ldots d\tau_n
\exp[-i\omega_1 \tau_1]\ldots \exp[-i\omega_n \tau_n]\Theta(\tau_1)\ldots
\Theta(\tau_n-\tau_{n-1})R^{(n)}(\tau_1,\ldots,\tau_n)\nonumber\\
&=\int_{-\infty}^\infty d\sigma_1 \ldots d\sigma_n \exp[-i\omega_1\sigma_1]
\exp[-i\omega_2\sum_{j=1}^2\sigma_j] \ldots \exp[-i\omega_n \sum_j^n
\sigma_j]\Theta(\sigma_1)\ldots \Theta(\sigma_n)S^{(n)}(\sigma_1,\ldots,\sigma_n)
\nonumber\\
&=\int_{-\infty}^\infty d\sigma_1 \ldots d\sigma_n \exp[-i\sigma_1\sum_{j=1}^n
\omega_j]\exp[-i\sigma_2\sum_{j=2}^n \omega_j] \ldots \exp[-i\omega_n
\sigma_n]\Theta(\sigma_1)\ldots \Theta(\sigma_n)S^{(n)}(\sigma_1,\ldots,\sigma_n)
\label{chisusc2},
\end{align}
where we have performed the change of variables  $\sigma_1=\tau_1$ and
$\sigma_j=\tau_j-\tau_{j-1}$ $\forall j\geq 2$ and we have defined
$S^{(n)}(\sigma_1,\ldots,\sigma_n)=R^{(n)}(\sigma_1,\ldots,\sum_j^n\sigma_j)$.
Defining $\Delta^{(n)}(\sigma_1,\ldots,\sigma_n)=\Theta(\sigma_1)\ldots
\Theta(\sigma_n)S^{(n)}(\sigma_1,\ldots,\sigma_n)$, we obtain that:
\begin{equation}
\chi^{(n)}(\omega_1,\ldots,\omega_n)=\hat{\Delta}^{(n)}(\sum_{j=1}^n\omega_j,\sum_{j=2}^n\omega_j,\ldots,\omega_n)\label{chidelta1},
\end{equation}
or, in other terms:
\begin{equation}
\chi^{(n)}(\nu_1-\nu_2,\nu_2-\nu_3,\ldots,\nu_n)=\hat{\Delta}^{(n)}(\nu_1,\nu_2,\ldots,\nu_n)\label{chidelta2}.
\end{equation}
Along the lines of the derivation proposed in the previous section, we
obtain:
\begin{align}
\hat{\Delta}^{(n)}(\omega_1,\ldots,\omega_n)& =\Pi_{i=1}^n
\left(-\frac{i}{\pi}\mathcal{P}\left(\frac{1}{\omega_{j_i}}\right)+\delta\left(\omega_{j_1}\right)\right)*\left(-\frac{1}{2^{k-1}}\Re\{\hat{\Delta}^{(n)}(\omega_1,\ldots,\omega_n)\}\right)\nonumber\\
&
=\Pi_{i=1}^n\left(-\frac{i}{\pi}\mathcal{P}\left(\frac{1}{\omega_{j_i}}\right)+\delta\left(\omega_{j_1}\right)\right)*\left(\frac{i}{2^{n-1}}\Im\{\hat{\Delta}^{(n)}(\omega_1,\ldots,\omega_n)\}\right)\nonumber\\
& =\Pi_{i=1}^n
\left(-\frac{i}{\pi}\mathcal{P}\left(\frac{1}{\omega_{j_i}}\right)+\delta\left(\omega_{j_1}\right)\right)*\left(\frac{1}{2^n}\hat{\Delta}^{(n)}(\omega_1,\ldots,,\omega_n)\right)\nonumber\\
& =\Pi_{i=1}^n
\left(-\frac{i}{\pi}\mathcal{P}\left(\frac{1}{\omega_{j_i}}\right)+\delta\left(\omega_{j_1}\right)\right)*\left(\frac{1}{2^n}\hat{S}^{(n)}(\omega_1,\ldots,,\omega_n)\right),
\label{kkgen2}
\end{align}
which highlights the fundamental connection, made possible by causality, between the
spectral properties of the response function and the corresponding susceptibility,
at all orders of nonlinearity. Equation \ref{kkgen2} suggests that the poles of
$\hat{\Delta}^{(n)}(\omega_1,\ldots,\omega_n)$ are, at most, those of
$\hat{S}^{(n)}(\omega_1,\ldots,\omega_n)$; moreover, the multiple convolution product ensures that all the singularities in the right hand side terms which are not compatible with causality are removed. 

Furthermore, if $S^{(n)}(\sigma_1,\ldots,\sigma_n)$ is even with respect to the
change of sign of all variables, so that $\hat{S}^{(n)}(\omega_1,\ldots,,\omega_n)$
is even and real, we have that the real part of
$\hat{\Delta}^{(n)}(\omega_1,\ldots,\omega_n)$ will be given by the sum of all
contributions in Eq. \ref{kkgen2} including an even number of convolutions between
$\hat{S}^{(n)}(\omega_1,\ldots,,\omega_n)$ and terms of the form
$\mathcal{P}\left(\frac{1}{\omega_{j_i}}\right)$, whereas the imaginary part,
conversely, will result from the remaining terms. Note that each time the
convolution product is applied, the parity of the function is exchanged. Same result
will hold if $S^{(n)}(\sigma_1,\ldots,\sigma_n)$ is odd: in this case, $i$ times the
imaginary part of the  $\hat{\Delta}^{(n)}(\omega_1,\ldots,\omega_n)$ will be given
by the sum of all contributions where  $\hat{S}^{(n)}(\omega_1,\ldots,,\omega_n)$
(which is odd and purely imaginary) is convolved an even number of times with the
factors $\mathcal{P}\left(\frac{1}{\omega_{j_i}}\right)$, and the real part of
$\hat{\Delta}^{(n)}(\omega_1,\ldots,\omega_n)$ will come from the remaining terms.

In the linear $n=1$ case, $\chi^{(1)}(\omega)=\hat{\Delta}^{(1)}(\omega)$ and the
results presented in Eqs. \ref{kkgen} and \ref{kkgen2} lead us to the classical
results presented by Kubo in the case of perturbations to Hamiltonian systems
immersed in a thermal bath \cite{Kubo}. In fact, we obtain that if $S^{(1)}(\tau_1)$ is even, it
is equal to $G_S^{(1)}(\tau_1)$, so that its Fourier Transform
$\hat{S}^{(1)}(\omega_1)$ is equal to $2\Re\{\chi^{(1)}(\omega)\}$,. Instead, if
$S^{(1)}(\tau_1)$ is odd, it is equal to $G_A^{(1)}(\tau_1)$, and we have that
$\hat{S}^{(1)}(\omega_1)=2i\Im\{\chi^{(1)}(\omega)\}$. Obviously, the parity properties of $S^{(1)}(\tau_1)$ depend critically on the unperturbed invariant measure and on the way the flow is perturbed, and so on the choice of $X(x)$. When nonlinear processes are
considered, the link between  $\hat{S}^{(n)}(\omega_1,\ldots,\omega_n)$ and
$\chi^{(n)}(\omega_1,\ldots,\omega_n)$ is indeed less trivial, even if Eqs.
\ref{chidelta2}-\ref{kkgen2} provide an algorithmically feasible way to unperturbed properties of the system to its response to external perturbations.

\section{Extending the FDT beyond the linear response: general treatment}
\label{sec:fdt1}
We now wish to explore how to link the (real or imaginary part) of the
susceptibility function at various orders of nonlinearity to the Fourier transform
of correlations of the system in the unperturbed state. In the linear case, this is
the fundamental content of the fluctuation-dissipation theorem. In order to pursue
this line, following Ruelle \cite{ruelle98}, we must assume that the unperturbed
invariant measure $\rho^{(0)}(dx)$ is absolutely continuous with respect to
Lebesgue, so that it can be expressed as
$\rho^{(0)}(dx)=\overline{\rho}^{(0)}(x)dx$. In this case, we can rewrite the linear
Green function given in Eq. \ref{Gn} for the case $n=1$ as a simple lagged
correlation between a function $C(x)$ and the observable at a later time
$A(x(\tau_1))$ evolved  according to the unperturbed dynamics: 
\begin{align}
G(\tau_1)& =\int dx \overline{\rho}^{(0)}(x) \Theta(\tau_1)X(x)\cdot\nabla
\Pi(\tau_{1})A(x)=-\int dx  \Theta(\tau_1) \overline{\rho}^{(0)}(x) \frac{\nabla
\cdot (\overline{\rho}^{(0)}(x) X(x))}{\overline{\rho}^{(0)}(x)} \Pi(\tau_{1})A(x)\\
& = \int dx  \Theta(\tau_1) \overline{\rho}^{(0)}(x) C(x) A(x(\tau_1))   = \int dx 
\Theta(\tau_1) \overline{\rho}^{(0)}(x) C(x(-\tau_1) A(x)  \label{G1c}.
\end{align}
because $\int dx (\nabla  \cdot (\overline{\rho}^{(0)}(x)
\Theta(\tau_1)X(x)\cdot\Pi(\tau_{1})A(x)))=0$, and where we have used the time
invariance of the measure $\overline{\rho}^{(0)}(x) dx$ in the last step of the
derivation. Note that Eq. \ref{G1c} provides a very general form of linear
fluctuation-dissipation theorem for dynamical systems endowed with a smooth invariant
measure.

In order to generalize this procedure for the $n^{th}$ order Green function, we
define the  adjoint operators for the operators $\Lambda$ and $\Pi(\sigma)$:
\begin{align}
\langle \alpha(x),\Lambda \beta(x) \rangle & = \langle \Lambda^+ \alpha(x) ,
\beta(x) \rangle \nonumber \\
\langle \alpha(x),\Pi(\sigma) \beta(x) \rangle & = \langle \Pi(\sigma)^+ \alpha(x) ,
\beta(x) \rangle 
\end{align}
where the scalar product $\langle\bullet,\bullet \rangle$ is the ordinary integral
evaluated on the support of $\overline{\rho}^{(0)}(x)$:
\begin{align}
\langle \alpha(x),\beta(x) \rangle  = \int_{\overline{\rho}^{(0)}(x)>0} dx
\alpha(x)\beta(x) 
\end{align}
Assuming that such support is compact or that the functions we consider vanish
sufficiently fast at infinity, we obtain that:   
\begin{align}
\Lambda (\beta(x)) =X(x)\cdot \nabla \beta(x) \rightarrow \Lambda^+(\alpha(x))
=-\nabla\cdot (X(x) \alpha(x)),
\end{align}
while the adjoint of the unperturbed evolution operator $\Pi(\tau_{1})$ is given by:  
\begin{align}
\Pi(\sigma)(\beta(x)) =\beta(x(\sigma)) \rightarrow \Pi(\sigma)^+(\alpha(x))
=\alpha(x(-\sigma)).
\end{align}
With these definitions, we derive formally from Eq. \ref{Gn} the following expression:
\begin{align}
R^{(n)}(\tau_1,\ldots,\tau_n)&=\langle \overline{\rho}^{(0)}(x),\Lambda
\Pi(\tau_n-\tau_{n-1}) \ldots \Lambda \Pi(\tau_{1})A(x)  \rangle\nonumber \\
&=\langle  \Pi(\tau_{1})^+  \Lambda^+  \ldots  \Pi(\tau_n-\tau_{n-1})^+  \Lambda^+
\overline{\rho}^{(0)}(x),A(x)\rangle,\label{Gnew}
\end{align}
which gives the $n^{th}$ order response function (and consequently, the Green
function)  as a $n-$ times correlation. Interestingly, when considering Eq.
\ref{Gnew}, one notes that the dual function $ \Pi(\tau_{1})^+  \Lambda^+  \ldots 
\Pi(\tau_n-\tau_{n-1})^+  \Lambda^+ \overline{\rho}^{(0)}(x)$ \emph{generates} the
Green functions corresponding to the perturbation flow $X(x)$ (given the unperturbed
variant measure $\overline{\rho}^{(0)}(x)$) for any considered observable,$A(x)$.
Furthermore, following from the definition given in Eq. \ref{chisusc2}, we obtain
the following expression for $S^{(n)}(\sigma_1,\ldots,\sigma_n)$:
\begin{align}
S^{(n)}(\sigma_1,\ldots,\sigma_n)&=\langle \overline{\rho}^{(0)}(x),\Lambda
\Pi(\sigma_n) \ldots \Lambda \Pi(\sigma_{1})A(x)  \rangle\nonumber \\
&=  \langle  \Pi(\sigma_{1})^+  \Lambda^+  \ldots  \Pi(\sigma_n)^+  \Lambda^+
\overline{\rho}^{(0)}(x),A(x)\rangle,
\end{align}

Combining  Eqs. \ref{gdef}- \ref{kkgen}, or, alternatively, Eqs.
\ref{chisusc2}-\ref{kkgen2} with the previous Eq. \ref{Gnew} and considering the
content of Eq. \ref{dissipa}, we obtain a generalized version of the FDT at all order of nonlinear and for rather general statistical
dynamical systems, specifically for those possessing a smooth invariant measure. In the next section we will show how to derive an actual explicit expression for the FDT in the special, albeit most relevant, case of the canonical ensemble.
  
\section{Extending the FDT beyond the linear response: Canonical Ensemble}
\label{sec:fdt2}
Following Kubo \cite{Kubo}, we now address explicitly the case of an interacting
many particle system whose unperturbed state is described by the canonical ensemble
generated by the Hamiltonian $H_0(x)$, which takes into account only the
\textit{internal} degrees of freedom, and analyze the impact of a adding a weak
perturbation Hamiltonian $H'(x,t)=B(x)T(t)$, where $B(x)$ is an observable
conjugated to the external field $T(t)$ \cite{Reichl,BLMV,ColRonVul}. In the Kubo
framework, the perturbed equations of motions can be written as 
$\dot{x}=F(x)+X(x)T(t)$, where $F(x)=S\cdot\nabla H_0(x)$ and $X(x)=S\cdot\nabla
B(x)$, where $S$ is the simplectic matrix.  Following the approach highlighted in
Ref. \cite{lucarini05}, one may adopt a perturbative technique to solve the Liouville
Equation for the probability density $\rho_t$. To this aim, one may formally write 
$$
\rho_t=\sum_{k=0}^{\infty}\rho^{(k)}_t \quad .
$$
This leads to the equation, valid for arbitrary order $n$:
\be
\frac{\partial \rho^{(n)}}{\partial t}=[H_0,\rho^{(n)}]+[B(x),\rho^{(n-1)}]T(t)
\label{rho}
\ee
where, in the classical case $[\bullet,\bullet]$ indicate the Poisson brackets
$\rho_{(0)}= \frac{e^{-\beta H_0}}{Z}$ denoting the (time-independent) equilibrium
canonical density, with $Z$ the canonical partition function, and where Eq.
(\ref{rho}) is supplemented with the initial condition $\rho_{t=0}=\rho_{(0)}$. In
the case of a quantum system, we can interpret $[\bullet,\bullet]$ as
$1/(i\hbar)\{\bullet,\bullet\}$, where $\{\bullet,\bullet\}$ is the canonical
commutator, and $\rho_{(0)}=\sum_a 1/Z\exp[-\beta E_a] |a\rangle \langle a|$ where
the $|a\rangle$'s constitute a complete set of eigenvectors of $H_0$.  As a result, one obtains that the expectation value of a given observable $A$ can be
written as \cite{zubarev}:
\begin{equation}
\langle A \rangle_t=\langle A \rangle_0+\sum_{n=1}^\infty \langle A \rangle^{(n)}_t,
\end{equation} 
where we revert to the notation $\langle \bullet \rangle_0 =Tr\{\rho_0
\bullet\}$, which is more common in the statistical physical literature, for indicating the expectation value $\rho_0(\bullet)$, in both the classical and quantum cases. The following expression
holds for the terms $n\geq 1$:
\be
\langle A \rangle^{(n)}_t = (-1)^n \int_{-\infty}^{\infty} d\tau_1 \ldots d\tau_n
\Theta(\tau_1)\Theta(\tau_2-\tau_1)\ldots \Theta(\tau_n-\tau_{n-1}) \left \langle
[B(-\tau_n),\ldots[B(-\tau_1),A]\ldots] \right\rangle_0 T(t-\tau_1)\ldots
T(t-\tau_n)\label{canonicalpert}
\ee
so that, following Eq. \ref{deltan}-\ref{Gn}, we can express the $n^{th}$ order
Green function as:
\be
G_{A,B}^{(n)}(\tau_1,\ldots,\tau_n)=(-1)^n\Theta(\tau_1)\Theta(\tau_2-\tau_1)\ldots
\Theta(\tau_n-\tau_{n-1}) \left \langle [B(-\tau_n),\ldots[B(-\tau_1),A]\ldots]
\right\rangle_0 
\ee
where the lower index of the Green function refers to the fact that we are
considering the perturbation to the observable $A$ due to the coupling with the $B$
field, whereas the response function $R_{A,B}^{(n)}$ is: 
\be
R_{A,B}^{(n)}(\tau_1,\ldots,\tau_n)=(-1)^n\left \langle
[B(-\tau_n),\ldots[B(-\tau_1),A]\ldots] \right\rangle_0 \label{start1} 
\ee
while the variable-wise rearranged function $S_{A,B}^{(n)}$ is: 
\be
S_{A,B}^{(n)}(\tau_1,\ldots,\tau_n)=(-1)^n\left \langle [B(-\sum_{j=1}^n
\tau_j),\ldots[B(-\tau_1),A]\ldots] \right\rangle_0 \label{start2}  
\ee
In the following, we will find a compact expression for its Fourier transform
$\hat{S}_{A,B}^{(n)}$, which, combined with what discussed in the previous sections,
provides the generalization of the FDT in the case of
perturbed Hamiltonian systems in contact with a thermostat at an inverse
temperature $\beta$.

\subsection{Equilibrium correlation functions: Linear case}
\label{sec:sec3}
\begin{figure}
   \centering
  \includegraphics[width=8.5cm]{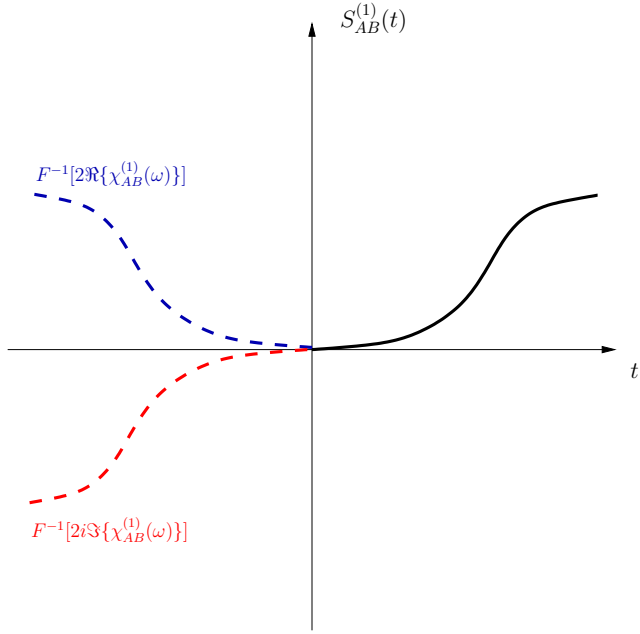}
   \caption{{Plot of the response function $S_{A,B}^{(1)}(t)$. Shown are the two cases when the latter is odd (red curve) and even (blue curve) under time reversal, which entails, respectively, $\hat{S}_{A,B}^{(1)}(\omega)=2i \Im\{\chi_{A,B}^{(1)})(\omega)\}$ and $\hat{S}_{A,B}^{(1)}(\omega)=2 \Re\{\chi_{A,B}^{(1)})(\omega)\}$.}}\label{evenodd}
\end{figure}

Let us now examine, order by order, how a general expression for the FDT emerges
from the previous results.  We first consider the linear response:
\bea
S_{A,B}^{(1)}(-t_1)&=R_{A,B}^{(1)}(-t_1)=&- \left \langle
[B(-t_1),A(0)]\right\rangle_0 \nonumber\\
&=& - \frac{1}{i\hbar}(\left \langle B(-t_1)A(0)\right\rangle_0-\left \langle A(0)
B(-t_1)\right\rangle_0)\nonumber\\ 
&=& - \frac{1}{i\hbar} (\mathcal{C}_{A,B}(-t_1)-\mathcal{C}_{A,B}(-t_1-\tau)) \label{1st}
\eea
with $\tau=i\hbar\beta$, and where $\mathcal{C}_{A,B}(t_1)=  Tr \{B(t_1)A(0)\rho^{(0)}\}$
denotes the two-time equilibrium correlation function. Moreover, in order to obtain
the last equality in (\ref{1st}), we employed the invariance of the trace under cyclic
permutations and the fact that the operator $e^{\beta H_0}$ effects a time
translation by the imaginary time $(-\tau)$.
Then, following Kubo \cite{Kubo}, when going to the Fourier space, it proves convenient to evaluate the complex conjugates of the various Fourier transforms.
Thus, for instance, one considers: 

\be
[\mathcal{F}(C_{A,B}(-t_1))]^{*}=[\hat{C}_{A,B}(-\omega_1)]^{*}=\mathcal{\hat{C}}_{A,B}(\omega_1)
\ee
where $\mathcal{\hat{C}}_{AB}(\omega_1)$ is the \textit{spectral density} \cite{Forster} and where the last equality follows from the the fact that $C_{AB}(-t_1)$ is real.
By Fourier transforming and by taking the complex conjugate on both sides of Eq. (\ref{1st}), we obtain:
\be
\hat{S}_{A,B}^{(1)}(\omega_1)=-\frac{1}{(i\hbar)}(1-e^{-\beta\hbar\omega_1})\mathcal{\hat{C}}_{A,B}(\omega_1)
\label{Delta1}
\ee
If, $S_{A,B}^{(1)}(t_1)$ is odd under time reversal, as it is commonly assumed, we have, as discussed in Sect. 2, that
$\hat{S}_{A,B}^{(1)}(\omega_1)=2i \Im\{\chi_{A,B}^{(1)})(\omega_1)\}$, so that:
\be
\Im\{\chi_{A,B}^{(1)})(\omega_1)\}=\frac{1}{2\hbar}(1-e^{-\beta\hbar\omega_1})
\hat{C}_{A,B}(\omega_1). \label{FDT1}
\ee
This represents the standard FDT at the first order in the perturbation. It connects the dissipation, related to $\Im\{\chi_{A,B}^{(1)})(\omega_1)\}$, to the
equilibrium fluctuations, contained in  $\hat{C}_{A,B}(\omega_1)$. On the other hand, if $S_{A,B}^{(1)}(t_1)$ is even under time reversal, it follows that
$\hat{S}_{A,B}^{(1)}(\omega_1)=2 \Re\{\chi_{A,B}^{(1)})(\omega_1)\}$, cf. Fig. \ref{evenodd}.

Starting from the general expression for the response function given in (\ref{start2}), which captures the physics of the problem, the strategy we employed at the linear order may be straightforwardly repeated at an arbitrary higher order, thus leading to a generalization of the FDT for deterministic systems.
For practical matters our expansion including second or third order around equilibrium is already new and relevant.

\subsection{Equilibrium correlation functions: second order}
We now present the first extension of the FDT beyond the linear order, thus showing
the explict calculations for the second order quantities. We start by writing out
explictly the expression for the response function:  
\bea
&&S_{A,B}^{(2)}(-t_1,-t_2)=R_{A,B}^{(2)}(-t_1,-t_1-t_2)= \langle
\frac{1}{(i\hbar)^2}[B(-t_1-t_2),[B(-t_1),A(0)]]\rangle  \nonumber\\
&=& \frac{1}{(i\hbar)^2}(\langle B(-t_1-t_2)B(-t_1)A(0)\rangle_0 - \langle B(-t_1-t_2)
A(0)
B(-t_1)\rangle_0 +\nonumber\\
&-& \langle B(-t_1)A(0)B(-t_1-t_2)\rangle_0 + \langle
A(0)B(-t_1)B(-t_1-t_2)\rangle_0)\nonumber\\
&=& \frac{1}{(i\hbar)^2}
(\mathcal{C}_{A,B}(-t_1,-t_1-t_2)-\mathcal{C}_{A,B}(-t_1-\tau,-t_1-t_2)-\mathcal{C}_{A,B}(-t_1,-t_1-t_2-\tau)+\mathcal{C}_{A,B}(-t_1-\tau,-t_1-t_2-\tau))
\label{2a}
\eea
where we have defined the correlation function
$\mathcal{C}_{A,B}(t_1,t_2)=\langle B(t_2)
B(t_1)A(0)\rangle_0$.
Then, we may consider, again, the complex conjugate of the Fourier
Transform of the three-time correlation functions occurring in Eq. (\ref{2a}), and find: 
\bea
\hat{\mathcal{C}}_{A,B}(\omega_1-\omega_2,\omega_2)&=&\int_{-\infty}^{+\infty}dt_2\int_{-\infty}^{+\infty}e^{-i(\omega_1
t_1+\omega_2
t_2)}\mathcal{C}_{A,B}(-t_1,-t_1-t_2) dt_1  \nonumber\\
&=&\int_{-\infty}^{+\infty}\int_{-\infty}^{+\infty}e^{i(\omega_1-\omega_2)\xi_1}e^{i\omega_2\xi_2}\mathcal{C}_{A,B}(\xi_1,\xi_2)d\xi_1
d\xi_2 \nonumber
\eea
with $\xi_1=-t_1$ and $\xi_2=-t_1-t_2$.
Thus, one obtains, from Eq.(\ref{2a}):
\bea
\hat{S}_{A,B}^{(2)}(\omega_1,\omega_2)&=\frac{1}{(-i\hbar)^2}(1-e^{-\beta\hbar\omega_2})(\hat{\mathcal{C}}_{A,B}(\omega_1-\omega_2,\omega_2)-e^{-\beta\hbar(\omega_1-\omega_2)}\hat{\mathcal{C}}_{A,B}(\omega_2,\omega_1-\omega_2))\nonumber
\\
&=\frac{1}{(-i\hbar)^2}(1-e^{-\beta\hbar\omega_2})\hat{\mathcal{C}}_{A,B}^{(2)}(\omega_1,\omega_2)
\label{Delta2}
\eea
where we have defined the generalized spectral density:
\bea
\hat{\mathcal{C}}_{A,B}^{(2)}(\omega_1,\omega_2):=\hat{\mathcal{C}}_{A,B}(\omega_1-\omega_2,\omega_2)-e^{-\beta\hbar(\omega_1-\omega_2)}\hat{\mathcal{C}}_{A,B}(\omega_2,\omega_1-\omega_2)
\quad . \label{Delta2bis}
\eea
Let us now exemplify how we can use these result to reconstruct the response of the
system starting from observing its fluctuations. We start by considering the
observable $B(-t_1-t_2)B(-t_1)A(0)$ and derive its expectation value $\mathcal{C}_{A,B}(-t_1,-t_1-t_2)=Tr \{B(-t_1-t_2)
B(-t_1)A(0)\rho^{(0)}\}$. We then compute the complex conjugate of the two-dimensional Fourier transform of
$\mathcal{C}_{A,B}(t_1,t_2)$ and obtain
$\hat{\mathcal{C}}_{A,B}(\omega_1-\omega_2,\omega_2)$. Using Eq.
\ref{Delta2bis}, we construct
$\hat{\mathcal{C}}_{A,B}^{(2)}(\omega_1,\omega_2)$ and, eventally, using
Eq. \ref{Delta2}, we obtain $S_{A,B}^{(2)}(\omega_1,\omega_2)$. Furthermore, we plug
$S_{A,B}^{(2)}(\omega_1,\omega_2)$ into Eq. \ref{kkgen2} and derive, via a double
convolution integral, the quantity $\Delta_{A,B}^{(2)}(\omega_1,\omega_2)$. Finally,
using the definition given in Eq. \ref{chidelta2}, we eventually obtain
$\chi_{A,B}^{(2)}(\omega_1,\omega_2)$, which contains the complete information on the
second order response of the system. Therefore, joining the statistical properties of
the fluctuations of the system to its response to external perturbations requires
linear changes of variables, simple algebraic sums and multiplications, and a
multiple convolution integral. These operations, albeit cumbersome, can be easily implemented
numerically. 

%

\subsection{Equilibrium correlation functions: third order}
We hereby present the explict calculations for the third order quantities. As easily
seen, the number of terms becomes almost unmanageable, but in the next subsection we
propose a general formula. We have:
\bea
&&S_{A,B}^{(3)}(-t_1,-t_2,-t_3)=R_{A,B}^{(3)}(-t_1,-t_1-t_2,-t_1-t_2-t_3)=\nonumber\\
&=&-\frac{1}{(i\hbar)^3}\langle[B(-t_1-t_2-t_3),[B(-t_1-t_2),[B(-t_1),A(0)]]]\rangle_0\nonumber\\
&=& -\frac{1}{(i\hbar)^3}
(\mathcal{C}_{A,B}(-t_1,-t_1-t_2,-t_1-t_2-t_3)-\mathcal{C}_{A,B}(-t_1-t_2,-t_1-t_2-t_3,-t_1-\tau)+
\nonumber\\
&-&
\mathcal{C}_{A,B}(-t_1,-t_1-t_2-t_3,-t_1-t_2-\tau)+\mathcal{C}_{A,B}(-t_1-t_2-t_3,-t_1-t_2-\tau,-t_1-\tau)+\nonumber\\
&-&
\mathcal{C}_{A,B}(-t_1,-t_1-t_2,-t_1-t_2-t_3-\tau)+\mathcal{C}_{A,B}(-t_1-t_2,-t_1-t_2-t_3-\tau,-t_1-\tau)+\nonumber\\
&+&
\mathcal{C}_{A,B}(-t_1,-t_1-t_2-t_3-\tau,-t_1-t_2-\tau)-\mathcal{C}_{A,B}(-t_1-t_2-t_3-\tau,-t_1-t_2-\tau,-t_1-\tau))
\label{3a}
\eea
where we have defined the four-time correlation function $\mathcal{C}_{A,B}(t_1,t_2,t_3)=Tr \{B(t_3) B(t_2)
B(t_1)A(0)\rho^{(0)}\}$.
Similarly to what obtained for the second order terms, we have that: 

\be
\hat{\mathcal{C}}_{A,B}(\omega_1-\omega_2,\omega_2-\omega_3,\omega_3)=\int_{-\infty}^{+\infty}\int_{-\infty}^{+\infty}\int_{-\infty}^{+\infty}e^{-i(\omega_1
t_1+\omega_2
t_2)}\mathcal{C}_{A,B}(-t_1,-t_1-t_2,-t_1-t_2-t_3) dt_1 dt_2 dt_3  
\ee
Thus, from Eq. (\ref{3a}) one finally obtains:
\bea
\hat{S}_{A,B}^{(3)}(\omega_1,\omega_2,\omega_3)&=&=-\frac{\epsilon_B^3}{(i\hbar)^3}(1-e^{-\beta\hbar\omega_3})(\mathcal{C}_{A,B}(\omega_1-\omega_2,\omega_2-\omega_3,\omega_3)-e^{-\beta\hbar(\omega_1-\omega_2)}\mathcal{C}_{A,B}(\omega_2-\omega_3,\omega_3,\omega_1-\omega_2)+\nonumber\\
&-&e^{-\beta\hbar(\omega_2-\omega_3)}\mathcal{C}_{A,B}(\omega_1-\omega_2,\omega_3,\omega_2-\omega_3)+e^{-\beta\hbar(\omega_1-\omega_2)}e^{-\beta\hbar(\omega_2-\omega_3)}\mathcal{C}_{A,B}(\omega_3,\omega_2-\omega_3,\omega_1-\omega_2))\nonumber\\
&=&
\frac{-1}{(-i\hbar)^3}(1-e^{\beta\hbar\omega_3})\hat{C}_{A,B}^{(3)}(\omega_1,\omega_2,\omega_3)
\label{Delta3}
\eea
where we have defined the function 
\bea
\mathcal{\hat{C}}_{A,B}^{(3)}(\omega_1,\omega_2,\omega_3)&:=&\mathcal{C}_{A,B}(\omega_1-\omega_2,\omega_2-\omega_3,\omega_3)-e^{-\beta\hbar(\omega_1-\omega_2)}\mathcal{C}_{A,B}(\omega_2-\omega_3,\omega_3,\omega_1-\omega_2)+\nonumber\\
&-&e^{-\beta\hbar(\omega_2-\omega_3)}\mathcal{C}_{A,B}(\omega_1-\omega_2,\omega_3,\omega_2-\omega_3)+e^{-\beta\hbar(\omega_1-\omega_2)}e^{-\beta\hbar(\omega_2-\omega_3)}\mathcal{C}_{A,B}(\omega_3,\omega_2-\omega_3,\omega_1-\omega_2)
\nonumber \quad .
\eea



\subsection{Equilibrium correlation functions: general formula}
The results obtained at the lower orders of the expansion pave the way for a straightforward generalization of the explicit formulae detailed above.
To this aim, let us define the $n$-time correlation function $\mathcal{C}_{A,B}(t_1,t_2,\ldots,t_n)=Tr \{B(t_n)\ldots
B(t_2)
B(t_1)A(0)\rho^{(0)}\}$. Then, by induction, it is possible to prove that, for an arbitrary
order $n$, the function $\mathcal{\hat{C}}_{A,B}^{(n)}(\omega_1,...,\omega_n)$ attains the
structure:

\bea
\mathcal{\hat{C}}_{A,B}^{(n)}(\omega_1,...,\omega_n) := \sum_{m=0}^{n-1}(-1)^m e^{-\beta \hbar
\sum_{k=1}^m \tilde{\omega}_{j_k}}
\sum_{j_m=m}^{n-1}\sum_{j_{m-1}=1}^{j_m-1}...\sum_{j_{1}=1}^{j_{2}-1} 
\mathcal{C}_{A,B}(\tilde{\omega}_1,...,\tilde{\omega}_n,\tilde{\omega}_{j_{m}},...,\tilde{\omega}_{j_1})
\nonumber\\ 
\label{general}  \quad .
\eea
with  
$$
\tilde{\omega}_k=\left\{
  \begin{array}{ll}
    \omega_k-\omega_{k+1}, & \hbox{for $k \in [1,n)$;} \\
    \omega_k, & \hbox{for $k = n$.}
  \end{array}\right. \quad .
$$  

For any $m\in[0,n-1]$, the sums on the r.h.s. of Eq.(\ref{general}) yield
$\frac{(n-1)!}{m!(n-m-1)!}$ terms (corresponding to all possible combinations of
time-ordered equilibrium correlation functions), and it is intended, in our
notation, that the term corresponding to $m=0$ yields
$\mathcal{\hat{C}}_{A,B}(\tilde{\omega}_1,...,\tilde{\omega}_n)$. 
The function  $\mathcal{\hat{C}}_{A,B}^{(n)}(\omega_1,...,\omega_n)$ generalizes, to an arbitrary order, the spectral density $\mathcal{\hat{C}}_{A,B}(\omega_1)$
appearing in Eq.(\ref{FDT1}).
In fact, the function $\hat{S}_{A,B}^{(n)}(\omega_1,...,\omega_n)$ is easily derived
from Eq. (\ref{general}) and reads:
\be
\hat{S}_{A,B}^{(n)}(\omega_1,...,\omega_n)=\frac{(-1)^n}{(i\hbar)^n}(1-e^{-\beta\hbar\tilde{\omega}_n})\mathcal{\hat{C}}_{A,B}^{(n)}(\omega_1,...,\omega_n)
\label{Deltagen}
\ee
As described in the subsection dedicated to the second order, it is possible to
define an experimental procedure for deducing the response of the system from the
spectral properties of the observable $\mathcal{C}_{A,B}(t_1,t_2,\ldots,t_n)$
thorugh a cumbersome yet straightforward set of operations.

\section{Conclusions}
\label{sec:conc}
The FDT represents a milestone in the endeavour towards a comprehensive theory aimed at connecting the internal fluctuations of a system to its response to external forcings. The seminal formulation proposed by Kubo \cite{Kubo} addressed linear deviations from equilibrium, and has found a vast range of applications in many fields of natural sciences. Recent investigations \cite{ruelle98,Sepulchre,majda07,ColRonVul} have tried to extend the FDT outside of the original domain by focusing on linear deviations from nonequilibrium steady states in the framework of of Axiom A dynamical systems. It has been underlined that when the unperturbed invariant measure is singular with respect to Lebesgue, as usually the case in deterministic  dissipative systems, a novel term arises, as shown by Ruelle \cite{ruelle98}, which seem to prevent immediate extension of the FDT around a non equilibrium steady state. This is related to the fact that for these systems forced and free fluctuations are not exactly equivalent \cite{ruelle09,lucarinisarno11}, yet recent works suggest the possibility of practically extending the range of validity of FDT also in this case \cite{ColRonVul} Nonetheless, a response theory can be formulated also in this case \cite{ruelle97,ruelle98,ruelle09}, and can be successfully framed in terms of frequency dependent susceptibilities at all order of nonlinearity \cite{lucarini08}.
In this work, instead, we partly pursue a different approach, with the goal of highlighting the strong link between causality and the possibility of connecting unambiguously fluctuation and response, both at linear and nonlinear level. We first show in a rather general setting how the formalism of the Ruelle response theory can be used to derive in a novel way Kramers-Kronig relations connecting the real and imaginary part of the the Fourier transforms of the $n$-th order Green function, i.e. the susceptibility $\chi^{(n)}(\omega_1,...,\omega_n)$. Moreover, we extend the Kramers-Kronig theory by showing that the application of multiple convolution integrals allow to derive the susceptibility from the Fourier transform of corresponding response function $S^{(n)}(\omega_1,...,\omega_n)$, cf. Sec.\ref{sec:respfun}. In this derivation, we shed light on the role of the causality principle (embodied by the sequence of $\theta$ functions forming the definition of the Green function, cf. Eq. (\ref{causal2}))  and of the time symmetries of the response function. Thus, Eqs. (\ref{chidelta1}), and (\ref{kkgen2}) represent a first, very general, result. Moreover, in the second part of the work, we focus on systems whose invariant measure is absolutely continuous with respect to Lebesgue and write a formal extension of the FDT to all orders of nonlinearity. We discuss in detail the case of a (classical or quantum) Hamiltonian system perturbed by an external field, described by the operator $B(t)$, from its equilibrium state, given in terms of the statistical density operator $\rho^{(0)}$. The idea underlying our approach was preliminarily proposed in \cite{EvMorr} in the context of classical Hamiltonian systems. Then, by resorting on a compact general expression available for the nonlinear Green function, we succeed to establish a link between the statistical properties of the \textit{equilibrium} fluctuations, incorporated in the generalized spectral density $\hat{C}_{A,B}^{(n)}(\omega_1,...,\omega_n)$ and the response, related to the function $\chi^{(n)}(\omega_1,\ldots,\omega_n)$ at an arbitrary order of nonlinearity.  In particular, we provide an exact expression for the generalized spectral density, Eq. (\ref{general}), which, supplemented with Eqs. (\ref{chidelta1}) and (\ref{kkgen2}), allows to establish a suitable extension of the FDT for nonlinear processes. While the FDT has an especially compact structure in the linear case, in the nonlinear case the derivation of the susceptibility of the system from the observation of suitably defined correlations  requires linear changes of variables, simple algebraic sums and multiplications, and a
multiple convolution integral. These operations are lengthy but overall trivial and of easy implementation. Concluding, in the Appendix we show the the imaginary part of the susceptibility at all orders is related to the power dissipated in the system if we select as observable the physical quantity conjugated to the external field. 
Our method also resembles the approach discussed in Ref. \cite{mattmaes}, where a compact stochastic version of a generalized FDT is accomplished by means of a perturbation theory around a reference equilibrium state, equipped with a detailed balance dynamics.
Further connections between the two perturbation theories, employed, respectively, in the deterministic and in the stochastic settings, would be worth investigating.

\appendix

\section{Dissipation and Susceptibility}
\label{sec:app}
It is possible to link the imaginary part of the susceptibility to the
time-integrated value of a quantity, which, in the case of Hamiltonian systems
in contact with a thermal bath and in the special context of linear processes, is related to the
energy dissipation (and consequent entropy production) of the system as a result of
the interaction with the perturbative field. Hereby, we wish to extend this result
in a more general context. We follow the formalism presented by Reichl
\cite{Reichl}. We then consider the quantity $P^{(n)}(t)=-T(t)\cdot d/dt
\rho^{(n)}_t(A)$, where $P(t)$ is a generalized power absorption of the system
associated to the interaction between the time varying field $T(t)$ and the conjugated
observable $A$. We can rewrite $P(t)$ as follows:
\begin{equation}
P^{(n)}(t)= -\frac{1}{(2\pi)^2} \int d\omega d\nu d\omega_1 \ldots
d\omega_1\hat{T}(\omega)\exp[-i\omega t] \frac{d}{dt} \exp[-i\nu
t]\chi^{(n)}(\omega_1,\ldots,\omega_n) \hat{T}(\omega_1)\ldots
\hat{T}(\omega_n)\delta(\nu-\sum_{j=1}^{n}\omega_j)
\end{equation}
\begin{equation}
P^{(n)}(t)= \frac{1}{(2\pi)^2} \int d\omega d\omega_1 \ldots d\omega_1
\hat{T}(\omega)\exp[-i\omega t] (i\sum_{j=1}^{n}\omega_j)
\exp[-i\sum_{j=1}^{n}\omega_j t]\chi^{(n)}(\omega_1,\ldots,\omega_n)
\hat{T}(\omega_1)\ldots \hat{T}(\omega_n).
\end{equation}
We integrate $P^{(n)}(t)$ over all times:
\begin{equation}
\int_{-\infty}^{\infty}dt P^{(n)}(t)=  \int  d\omega_1 \ldots d\omega_1
\hat{T}(-\sum_{j=1}^{n}\omega_j) (i\sum_{j=1}^{n}\omega_j)
\chi^{(n)}(\omega_1,\ldots,\omega_n) \hat{T}(\omega_1)\ldots \hat{T}(\omega_n),
\end{equation}
and derive that the total power absorption at order $n$ comes from coupling the
response at frequency $i\sum_{j=1}^{n}\omega_j$ with the incoming field at the
opposite frequency. Therefore, the larger the band of the time modulation $T(t)$,
the easier it will be to find matching conditions on the frequency. Since $T(t)$ is
a real function, we have that $T(\omega)=(T(\omega))^*$. Therefore, since $P(t)$ is
a real function, we derive that:    
\begin{equation}
\int_{-\infty}^{\infty}dt P^{(n)}(t)= - \int  d\omega_1 \ldots d\omega_1
(\sum_{j=1}^{n}\omega_j)\Im\{ \chi^{(n)}(\omega_1,\ldots,\omega_n)\}
\hat{T}(\omega_1)\ldots
\hat{T}(\omega_n)\hat{T}(-\sum_{j=1}^{n}\omega_j)\label{dissipa} ,
\end{equation}
which proves that the at all order of nonlinearity, the imaginary part of the
susceptibility describes the power dissipation of the system as defined by
$P^{(n)}(t)=-T(t)\cdot d/dt \rho^{(n)}_t(A)$. In the special case of impulsive
perturbations, so that $T(t)=T_0\delta(t)\rightarrow \hat{T}(\omega)=T_0$, we have:
\begin{equation}
\int_{-\infty}^{\infty}dt P^{(n)}(t)= - \int  d\omega_1 \ldots d\omega_1
(\sum_{j=1}^{n}\omega_j)\Im\{ \chi^{(n)}(\omega_1,\ldots,\omega_n)\} T_0^{n+1}.
\end{equation}
One must note that in the case of a non-continuum spectrum time modulation $T(t)$ -
e.g. when $T(t)$ is constituted by $2m$ frequency components (positive and negative)
- contributions to the absorption at $n^{th}$ order will come only from the generate
terms where the sum of $n$ frequencies chosen, possibly with repetition, among the
$2m$ frequencies of $T(t)$ match one of the $2m$ frequencies of $T(t)$ itself. In
particular, in the case of a monochromatic input $T(t)$, no absorption will take
place, e.g., at all even orders of nonlinearity.

\end{document}